\documentclass[11pt]{article}
\usepackage{cite}
\usepackage{amsmath,amssymb,amsfonts}
\usepackage{graphicx}
\usepackage{textcomp}
\usepackage{xcolor}
\usepackage{parskip}
\usepackage{float,epsfig}
\usepackage{color}
\usepackage{cancel}
\usepackage{rotating}
\usepackage{lscape}
\usepackage{braket}
\usepackage{subcaption}
\usepackage{tikz}
\usepackage{booktabs}
\usepackage{todonotes}
\usepackage{algpseudocode}
\usepackage{algorithm}
\usepackage{hyperref}
\usepackage{cleveref}
\usepackage{comment}
\usepackage{qcircuit}
\usepackage{xcolor}

\usepackage{lipsum}



\begin{document}

\newtheorem{theorem}{\bf Theorem}[section]
\newtheorem{proposition}[theorem]{\bf Proposition}
\newtheorem{corollary}[theorem]{\bf Corollary}
\newtheorem{lemma}[theorem]{\bf Lemma}

\newtheorem{definition}[theorem]{\bf Definition}
\newtheorem{example}[theorem]{\bf Example}
\newtheorem{exam}[theorem]{\bf Example}

\newtheorem{remark}[theorem]{\bf Remark}
\newtheorem{observation}[theorem]{\bf Observation}
\newcommand{\nrm}[1]{|\!|\!| {#1} |\!|\!|}

\newcommand{\ba}{\begin{array}}
\newcommand{\ea}{\end{array}}
\newcommand{\von}{\vskip 1ex}
\newcommand{\vone}{\vskip 2ex}
\newcommand{\vtwo}{\vskip 4ex}
\newcommand{\dm}[1]{ {\displaystyle{#1} } }

\newcommand\independent{\protect\mathpalette{\protect\independenT}{\perp}}
\def\independenT#1#2{\mathrel{\rlap{$#1#2$}\mkern2mu{#1#2}}}
\def \pf{{\bf Proof: }}

\newcommand{\be}{\begin{equation}}
\newcommand{\ee}{\end{equation}}
\newcommand{\beano}{\begin{eqnarray*}}
\newcommand{\eeano}{\end{eqnarray*}}
\newcommand{\inp}[2]{\langle {#1} ,\,{#2} \rangle}
\def\bmatrix#1{\left[ \begin{matrix} #1 \end{matrix} \right]}
\def\dmatrix#1{\left| \begin{matrix} #1 \end{matrix} \right|}
\def \noin{\noindent}
\newcommand{\evenindex}{\Pi_e}


\def \N{{\mathbb N}}
\def \Z{{\mathbb Z}}
\def \R{{\mathbb R}}
\def \C{{\mathbb C}}
\def \K{{\mathbb K}}
\def \J{{\mathcal J}}
\def \Q{{\mathbb Q}}
\def \Uf{{\mathsf U}}
\def \Sf{{\mathsf S}}
\def \sf{{\mathfrak s}}
\def \uf{{\mathfrak u}}
\def \calL{\mathcal{L}}

\def \calB{{B}}
\def \calK{\mathcal{K}}
\def \calC{\mathcal{C}}
\def \calP{\mathcal{P}}
\def \calD{{D}}
\def \calV{{V}}
\def \calG{{G}}
\def \calN{\mathcal{N}}
\def \calT{\mathcal{T}}
\def \calH{\mathcal{H}}
\def \calM{\mathcal{M}}
\def \calI{\mathcal{I}}
\def \calE{{E}}
\def \calU{{U}}
\def \norm{\nrm{\cdot}\equiv \nrm{\cdot}}

\def \diag{\mbox{diag}}
\def \tr{\mathrm{Tr}}
\def \lam{\lambda}
\def \sig{\sigma}
\def \Sig{\Sigma}
\def \Lam{\Lambda}
\def \ep{\epsilon}
\def \sgn{\mathrm{sgn}}
\def \det{\mathrm{det}}

\algrenewcommand\algorithmicrequire{\textbf{Input:}}
\algrenewcommand\algorithmicensure{\textbf{Output:}}
\newcommand{\cupdot}{\mathbin{\mathaccent\cdot\cup}}

\def\BibTeX{{\rm B\kern-.05em{\sc i\kern-.025em b}\kern-.08em
    T\kern-.1667em\lower.7ex\hbox{E}\kern-.125emX}}

\newcommand{\tm}[1]{\textcolor{magenta}{ #1}}
\newcommand{\tre}[1]{\textcolor{red}{ #1}}
\newcommand{\tb}[1]{\textcolor{blue}{ #1}}


\title{Scalable quantum circuits for $n$-qubit unitary matrices\\
}

\author{Rohit Sarma Sarkar\thanks{
Department of Mathematics, IIT Kharagpur, India, Email: 
rohit15sarkar@yahoo.com} \,\,\,\,\,\,\, 
Bibhas Adhikari\thanks{Department of Mathematics, IIT Kharagpur, India, Email:
bibhas@maths.iitkgp.ac.in}
}

\date{}

\maketitle
\thispagestyle{empty}

\noindent{\bf Abstract}
This work presents an optimization-based scalable quantum neural network framework for approximating $n$-qubit unitaries through generic parametric representation of unitaries, which are obtained as product of exponential of elements of a new basis that we propose as an alternative to Pauli string basis. We call this basis as the Standard Recursive Block Basis, which is constructed using a recursive method, and its elements are permutation-similar to block Hermitian unitary matrices.\\


\section{Introduction}
Parametric representation of unitary matrices has been an active area of research for a long time. With the recent advancements in quantum hardware technology, this has become even more crucial in the field of quantum computing since quantum evolution is described by unitary matrices. Additionally, the emergence of NISQ computers and variational quantum algorithms necessitates the implementation of quantum processes through parametrized quantum circuits consisting of only a few universal quantum gates. Thus it is of paramount interest to design quantum parametrized circuits for unitary evolution on a multi-qubit system using one-qubit and two-qubit quantum gates \cite{sete2016functional}.   

The Solovay-Kitaev theorem, established by Kitaev and Solovay, proves that any one-qubit special unitary matrix can be approximated by a collection of one-qubit gates, which can generate a dense subset of the $\Sf\Uf(2)$ \cite{kitaev1997quantum}. This process of identifying suitable approximations for a given unitary matrix is known as ``compilation" for fault-tolerant quantum computation \cite{paler2017fault,harrow2001quantum}. It is well-known that $CNOT$ and one-qubit gates form a universal model of quantum computation and can represent unitaries for multi-qubit systems \cite{harrow2001quantum,barenco1995elementary,zhou2000methodology,matsumoto2008representation,wang2011surface,PhysRevA.87.042302,bocharov2015efficient,ross2016optimal,heyfron2018efficient,venturelli2018compiling}. 

In the NISQ era, several proposals are made for physical implementations of near-term quantum computers. The compilation problem for decomposing a given unitary in terms of the primitive gate set of a near-term quantum device and approximating the unitary with high accuracy is one of the fundamental among them. There have been advancements in efficiently solving the quantum compilation problem using various methods such as recursive CS decomposition and Quantum Shannon-decomposition \cite{mottonen2004quantum,krol2022efficient}. 

Recently, an optimization based viewpoint for the compilation problem has generated a lot of interest \cite{madden2022best,nakajima2005new}.  In this approach, a unitary matrix is found that can be realized in hardware with constraints (such as the native gate alphabets of the NISQ platform), that is the closest to a target unitary with respect to a metric. Various cost functions are defined in these optimization-based approaches to achieve a good implementation of the target unitary. For example, optimizing the structure (i.e., where to place a $CNOT$ gate), optimizing the rotation angles of the rotation gates, optimizing the number of $CNOT$ count etc. after writing a parametric representation using matrix decomposition of the target unitary \cite{shende2005synthesis,shende2004minimal,bilek2022recursive}. 

In this paper, we present an optimization-based approach to approximate a given unitary matrix through generic parametrized unitary matrices. This leads to the development of a quantum neural network framework for implementing $n$-qubit unitaries using quantum circuits of $CNOT$ and one-qubit gates. To obtain a generic parametrized representation for unitaries, a new basis for the algebra of $d \times d$ complex matrices is introduced, with the aim of expressing any unitary through product of exponential of the proposed basis elements. We call this basis as Recursive Block Basis (RBB) and the basis elements are Hermitian and unitary matrices that are either diagonal or $2$-sparse matrices, alike the Pauli string basis elements. The proposed basis has an advantage over the Pauli string basis as the basis elements are permutation similar to block diagonal matrices, making it easier to compute the exponential of the basis elements. Further, we use this basis to define a basis for $n$-qubit systems with trace-zero matrices that we call Standard Recursive Block Basis (SRBB), which contains $2^n$ diagonal basis elements that are Pauli strings with Pauli matrices $I_2$ and $\sigma_3,$ the 
Pauli basis matrix about $Z$ axis. 

We present an algorithm for approximating any target unitary of order $d$ using generic parametrized unitary matrices with an ordering of exponential of the basis elements for multiplying to approximate a target unitary such that the number of $CNOT$ gate is reduced for a quantum circuit representation of the algorithm. The optimal values of the parameters to determine a unitary close to a target unitary are obtained using a classical optimization algorithm, such as Nelder-Mead, to minimize the Frobenius norm of the distance between the target and generic parametrized unitary matrices. The quantum circuit formulation of generic unitary matrices for $n$-qubit systems is defined by a quantum neural network framework with $L\geq 1$ layers. The recursive approach used to construct our basis also has the advantage that the proposed quantum circuit for $n$-qubit systems is scalable. Given a circuit for $n$-qubits, the circuit for $(n+1)$-qubits can be implemented using the current circuit with the addition of new $CNOT$ gates and $1$-qubit rotation gates. The proposed quantum circuit of one layer of approximation has at most $2 \cdot 4^n + (n-5)2^n$ $CNOT$ gates, and at most $\frac{3}{2} \cdot 4^n - \frac{5}{2} \cdot 2^n +1$ one-qubit rotation gates corresponding to $Z$ axis. We examine various scenarios to evaluate the effectiveness of our approximation algorithm in approximating standard and random unitary matrices for $2$-qubit, $3$-qubit, and $4$-qubit systems. Our results indicate that the proposed algorithm performs better one layer of approximation when the target unitaries are sparse, and the error of the approximation reduces with the increase of number of layers for the approximation. Lastly, we present an algorithm that enables the implementation of the proposed quantum circuits from $n$-qubit to $(n+1)$-qubit systems.


The remainder of the paper is structured as follows. In Section \ref{Sec:3}, recursive methods are introduced for constructing a basis of Hermitian, unitary matrices for complex matrices of size $d\times d$. Section \ref{Sec:4} presents a quantum neural network framework for an optimization-based approximation algorithm to approximate target unitaries for $n$-qubit systems. In Section \ref{sec:circuit}, a scalable quantum circuit is proposed for approximation algorithm. 

\section{Recursive approaches for the construction of a Hermitian unitary basis}\label{Sec:3}


The following theorem describes a recursive approach for construction of a Hermitian unitary basis, with $d^2-1$ of them having trace zero when $d$ is even. We call this basis as the {\bf R}ecursive {\bf B}lock {\bf B}asis (RBB).

\begin{theorem}\label{thm:basis2} ({\bf RBB})
Let $\boldsymbol{\mathcal{B}^{(d)}}=\{B^{(d)}_j : 1\leq j\leq d^2\},$ $d> 2$ denote the desired ordered basis for the matrix algebra of $d\times d$ complex matrices. Then setting $\boldsymbol{\mathcal{B}^{(2)}}=\{\sigma_1.\sigma_2,\sigma_3,I_2\}$ as the Pauli basis, the elements of $\boldsymbol{\mathcal{B}^{(d)}}$ can be constructed from the elements of $\boldsymbol{\mathcal{B}^{(d-1)}}$ using the following recursive procedure: \begin{small}$$B^{(d)}_j=
     \begin{cases}
     \left[ 
    \begin{array}{c|c} 
     B^{(d-1)}_{j}  & 0 \\ 
      \hline 
      0 & (-1)^{d-1}
    \end{array} 
    \right]; \,\, \mbox{if} \,\, j \in \{1,\hdots,(d-1)^2-1\}, \\
      
      P_{(k,d-1)}\left[ 
    \begin{array}{c|c} 
      D & 0 \\ 
      \hline 
      0 & \sigma_1
    \end{array} 
    \right]  P_{(k,d-1)}; \\ \mbox{if} \,\, j=(d-1)^2+(k-1),  \\ 
       
      P_{(k,d-1)}\left[ 
    \begin{array}{c|c} 
      D & 0 \\ 
      \hline 
      0 & \sigma_2 
    \end{array} 
    \right]P_{(k,d-1)};  \\ \mbox{if} \,\, j=(d-1)^2+(d-1)+(k-1),\\ 
     
     \left[ 
    \begin{array}{c|c} 
      I_{\lfloor d/2\rfloor +1} & 0 \\ 
      \hline 
      0 & -I_{\lfloor d/2\rfloor}
    \end{array} 
    \right]; \,\, \mbox{if} \,\, j = d^2-1 \,\, \mbox{and} \,\, d \,\, \mbox{is odd}\\
    
     \left[ 
    \begin{array}{c|c} 
      \Sigma & 0 \\ 
      \hline 
      0 & \sigma_3  \\
    \end{array} 
    \right]; \,\, \mbox{if} \,\, j = d^2-1 \,\, \mbox{and} \,\, d \,\, \mbox{is even} \\
    
    I_d \,\, \mbox{if} \,\, j = d^2 
     \end{cases},$$ \end{small}
     where  $k\in\{1,\hdots,d-1\}$ $P_{k,(d-1)}$ is the permutation matrix of order $d$ corresponding to the $2$-cycle $(k,d-1),$ $D=\mbox{diag}\{d_l : 1\leq l\leq d-2\},$ $d_l=(-1)^{l-1},$ and $\Sigma=\bmatrix{I_{\lfloor d/2\rfloor-1} & 0 \\ 0 & -I_{\lfloor d/2\rfloor-1}}$ Besides, $$\tr(B_j^{(d)})=\begin{cases}
      1 \,\, \mbox{if} \,\, d \,\, \mbox{is odd} \\
      0 \,\, \mbox{if} \,\, d \,\, \mbox{is even,}
     \end{cases},$$ $1\leq j\leq d^2-1,$ $\left(B_j^{(d)}\right)^2=I_d,$ and $\{B_j^{(d)} : 1\leq j\leq d^2-1\}$ forms a basis for $su(d)$ when $d$ is even. The basis elements that are diagonal matrices are given by $B^{(d)}_j$ where $j=m^2-1, 2\leq m\leq d$ and $B^{(d)}_{d^2}=I_d.$
 \end{theorem}
 
 \pf First observe that the matrices $B^{(d)}_j, 1\leq j\leq d^2$ are Hermitian and unitary due to the construction. Also, $\tr(B_j^{(d)})=0$ when $d$ is even and $\tr(B_j^{(d)})=1$ when $d$ is odd. Now we show that these matrices form a linearly independent subset of $\C^{d\times d}.$ Suppose $d$ is even. Then set
 \begin{small}
 \begin{eqnarray}0 &=& 
 \sum_{m=1}^{(d-1)-1^2} \underbrace{\bmatrix{c_{1m}B_m^{(d-1)}&0\\0&-c_{1m}}}_A + \nonumber \\ 
 && \sum_{m=1}^{(d-1)} \underbrace{P_{(m(d-1))} \bmatrix{(c_{2m}+c_{3m})D &0\\0&c_{2m}\sigma_1+c_{3m}\sigma_2} P_{(m(d-1))}}_B \nonumber \\ 
 && + \underbrace{\bmatrix{c_{44}\Sigma +c_{55}I_{d-2}&0\\0& -c_{44}\sigma_3+c_{55}I_2}}_C. \label{eqn:pf1} \end{eqnarray}
 \end{small}

Then see that the first $d-1$ entries of the last column of $B$ are given by $c_{2m}-ic_{3m},$ $1\leq m\leq d-1,$ whereas these corresponding entries in $A$ and $C$ are zero. Also first $n-1$ entries (left to right) of the last row of $B$ are given by $c_{2m}+ic_{3m},$ $1\leq m\leq d-1,$ whereas these corresponding entries in $A$ and $C$ are zero. Then equating the last column and row of the rhs of equation (\ref{eqn:pf1}) with the last zero column and row of lhs, it follows that $c_{2m}=c_{3m}=0,$ $1\leq m\leq d-1.$ Then the equation (\ref{eqn:pf1}) becomes \begin{eqnarray} 0 &=& \sum_{m=1}^{(d-1)^2-1} \bmatrix{c_{1m}B_m^{(d-1)}&0\\0&-c_{1m}} + \nonumber \\ &&{\bmatrix{c_{44}\Sigma +c_{55}I_{d-2}&0\\0& -c_{44}\sigma_3+c_{55}I_2}}. \end{eqnarray} Further, since $\{B_m^{(d-1)} : 1\leq m\leq (d-1)^2-1\}\cup I_{d-1}$ is linearly independent, then using the same method described above, the matrix $\sum_{m=1}^{(d-1)^2-1}c_{1m}B_j^{(d-1)}$ has all non-diagonal entries $0$. Thus the only terms remain are diagonal matrices i.e. the equation reduces to \begin{small} \begin{eqnarray} 0 &=&\sum_{m=2}^{(d-1)} \bmatrix{c_{1(m^2-1)}B_{(m^2-1)}^{(d-1)}&0\\0&-c_{1(m^2-1)}} + \nonumber \\ 
&& {\bmatrix{c_{44}\Sigma +c_{55}I_{d-2}&0\\0& -c_{44}\sigma_3+c_{55}I_2}} \label{eqn:LI3}  
  \end{eqnarray} \end{small} where $B^{(d-1)}_{m^2-1}$ and $I_{d-1}=B^{(d-1)}_{(d-1)^2},$ $2\leq m\leq d-1$ are proposed basis elements of $\C^{(d-1)\times (d-1)}.$ 
 
 For a diagonal matrix $M$ of order $d$ with diagonal entries $m_{jj}, 1\leq j\leq d,$ set $\diag(M)=[m_{11} \, m_{22} \, \hdots \, m_{dd}]^T$ as the column vector. Then observe that equation (\ref{eqn:LI3}) can be described as a linear system $Ax=0,$ where $x=\bmatrix{c_{13}& \hdots & c_{1((d-1)^2-1)} & c_{44} & c_{55}}^T$ and the consecutive columns of $A$ are 
given by $\diag\left(B^{(d)}_{m^2-1}\right),$ $2\leq m\leq d-1,$ $\diag\left(\bmatrix{\Sigma&0&0\\ 0&-1&0\\ 0&0&1}\right),$ and $\diag\left(I_{d}\right).$

 
 Next we show that $A$ is non-singular i.e. the columns of $A$ form a linearly independent set. Suppose \begin{small}\begin{eqnarray} 0 &=& \sum_{m=2}^{d-1} \alpha_m\bmatrix{\diag\left(B^{(d)}_{m^2-1}\right) \\ -1} + \nonumber \\ 
 && \beta \bmatrix{\diag\left(\bmatrix{\Sigma&0&0\\ 0&-1&0 \\ 0&0&1}\right)}  + \gamma \bmatrix{\diag\left(I_{d}\right)}. \nonumber\end{eqnarray} \end{small} Then multiplying the all-one vector ${\bf 1}_d^T$ from left at the above equation, we obtain $n\gamma =0$ since sum of entries of all other vectors are zero. This further implies $\gamma=0.$ Thus we have \begin{small}$$\sum_{m=2}^{d-1} \alpha_m\bmatrix{\diag\left(B^{(d)}_{m^2-1}\right) \\ -1}  + \beta \bmatrix{\diag\left(\bmatrix{\Sigma&0&0\\ 0&-1&0 \\ 0&0&1}\right)}=0.$$\end{small}
 Now note that the first entry of all the vectors in the above vectors are $1.$ Then considering the first and last entries of the above vectors, we obtain \begin{small}$$
 \beta+\sum_{m=2}^{d-1} \alpha_m = 0 \,\, \mbox{and} \,\,
 \beta -\sum_{m=2}^{d-1} \alpha_m = 0,$$\end{small} whose only solution is $\beta=\alpha_m=0$ for all $m.$ Hence the desired result follows when $m$ is even. The proof for odd $m$ follows similarly.  \hfill{$\square$}

\begin{remark}
 
   \begin{itemize}
       \item[(a)] Note that any of the  basis elements described by the above theorem that is a non diagonal matrix, is one of the form $$P\bmatrix{D_1 & 0 & 0 \\ 0 & \sigma & 0 \\ 0&0& D_2}P, \,\, \,\, P\bmatrix{\sigma & 0\\ 0 & D}P, \,\,\,\, P\bmatrix{D &0 \\0 & \sigma}P$$ where $D, D_1, D_2$ are diagonal matrices with entries from $\{1,-1\}$, $\sigma\in\{\sigma_1,\sigma_2\}$ and $P$ is a $2$-cycle or transposition. 
       \item[(b)] The basis elements with indices $j\in\mathcal{J}=\{l^2-1 : 2\leq l\leq d\}\cup\{d^2\}$ are diagonal matrices, which are orthogonal to each other. Obviously, $|\mathcal{J}|=d-1.$
   \end{itemize}  
\end{remark}

Now we present a Hermitian unitary trace-less basis of $n$-qubit systems that will play a crucial role in the remainder of the paper. The idea is that we now replace the diagonal basis elements of $\boldsymbol{\mathcal{B}^{(2^n)}}$ described in Theorem \ref{thm:basis2} by another set of diagonal matrices keeping invariance of the linearly independent property of the basis. First note that the set of matrices \begin{equation}\label{eqn:diag}\mathcal{D}_{IZ}=\{A_1\otimes \hdots \otimes A_n : A_j\in\{I_2,\sigma_3\}, 1\leq j\leq n\}\end{equation} is a set of $2^n$ linearly independent diagonal matrices with trace zero except when $A_j=I_2$ for all $j$ i.e. $A_1\otimes \hdots\otimes A_n=I_{2^n}.$ We call the proposed basis as the {\bf S}tandard {\bf R}ecursive {\bf B}lock {\bf B}asis (SRBB), defined below.

\begin{corollary}\label{cor:basis} ({\bf SRBB})
Let $\boldsymbol{\mathcal{B}^{(2^n)}}=\{B^{({2^n})}_j : 1\leq j\leq {2^{2n}}\}$ denote the basis described in Theorem \ref{thm:basis2}, and $\mathcal{D}_{IZ}$ is given by equation (\ref{eqn:diag}). Then the set $\boldsymbol{\mathcal{U}}^{(2^n)}=\{U^{(2^n)}_j : 1\leq j\leq 2^{2n}\}$, where $$U^{(2^n)}_j = \begin{cases}
 D \in\mathcal{D}_{IZ}\,\, \mbox{if} \,\, j\in \mathcal{J}=\{l^2-1 : 2\leq l\leq 2^n\}\cup\{2^{2n}\} \\
 B_j^{(2^n)}, \,\, \mbox{otherwise} 
\end{cases}$$ forms a Hermitian unitary basis for $\C^{2^n\times 2^n}$. Besides, $\tr(U_j^{(2^n)})=0$ when $U^{(2^n)}_j\neq I_{2^n}.$     
\end{corollary}

Now we introduce a function which provides an ordering of the diagonal elements of $\boldsymbol{\mathcal{U}}^{(2^n)}.$ From now onward, we denote $A_1\otimes A_2\otimes \hdots\otimes A_m=\otimes_{i=1}^m A_i$ for some matrices or vectors $A_i.$ If $A_i=A$ for all $i$ then we denote $\otimes_{i=1}^m A_i=\otimes^m A.$ 

 \begin{definition}\label{definition2}
 Define $\chi:\{I,Z\}\rightarrow \{0,1\}$ such that $\chi{(I)}=0,\chi(Z)=1.$ For any positive integer $m,$ define $\chi_m:\{\otimes_{i=1}^m A_i \,|\, A_i\in \{I,Z\},1\leq i\leq m\}\rightarrow \{0,1,\hdots,2^m-1\}$ such that $$\chi_m\left(\otimes_{i=1}^m A_i\right)=\sum_{i=1}^m 2^{i-1}\chi(A_i).$$
 \end{definition}


\section{Parametric representation of unitary matrices}\label{Sec:4}
It is well-known that the set of all unitary matrices of order $d,$ denoted by $\Uf(d)$ forms a Lie group and the corresponding Lie algebra is the real vector space of all skew-Hermitian matrices of order $d$ which we denote as $\uf(d).$ A classification of unitary matrices is that: any unitary matrix can be expressed as exponential of a skew-Hermitian matrix i.e. the map $\exp: \uf(d)\rightarrow \Uf(d)$ is surjective [Theorem 3.2, \cite{gallier2020differential}]. Now we develop a parametric representation of  unitary matrices of order $d.$

We recall from [Chapter 2, \cite{varadarajan2013lie}] that if $\{X_1,\hdots, X_k\}$ is a basis of the Lie algebra of a Lie group $G$ then for some $\theta>0,$ the map \begin{small}$$\psi : (\theta_1,\theta_2, \hdots, \theta_k) \mapsto \exp(\theta_1X_1)\exp(\theta_2X_2)\hdots\exp(\theta_kX_k)$$\end{small} from $\R^k$ into $G$ is an analytic diffeomorphism of the cube $I_\theta^k=\{(\theta_1,\hdots,\theta_k) : |\theta_j|<\theta, 1\leq j\leq k\}$ of $\R^k$ onto an open subset $U$ of $G$ containing the identity element $I$ of $G.$ If $x_1,\hdots,x_k$ are the analytic functions on $U$ such that the map $y\mapsto (x_1(y),\hdots, x_k(y))$ inverts $\psi,$ then for $1\leq j\leq k,$ \begin{small}$$x_j(\exp \theta_1X_1, \exp \theta_2X_2, \hdots,\exp \theta_kX_k)=\theta_j, (\theta_1,\hdots,\theta_k)\in I^k_\theta.$$\end{small} Then $x_1,\hdots,x_k$ are called the canonical coordinates of the second kind around $I$ with respect to the basis $\{X_1,\hdots, X_k\}.$ 

Setting $G=\Uf(d)$, the dimension of $\uf(d)$ is $d^2$ and if $\{B_j^{(d)} : 1\leq j\leq d^2\}$ denotes a basis of $\uf(d)$ then we have the following theorem. 

\begin{theorem}\label{thm:span}
There exists a $\theta>0$ such that $\left\{\prod_{j=1}^{d^2} \exp\left(i\theta_jB_j^{(d)}\right) : (\theta_1,\hdots,\theta_{d^2}) \in I^{d^2}_\theta \right\}$ generates $\Uf(d).$ 
\end{theorem}
\pf With the standard subspace topology of the matrix algebra of complex matrices, $\Uf(d)$ is a connected topological space. Then there exists $\theta>0$ such that the map $\psi:  (\theta_1,\hdots,\theta_{d^2})\mapsto \exp(\theta_1B_1^{(d)}),\hdots, \exp(\theta_{d^2}B^{(d)}_{d^2})$ is a diffeomorphism from $I^{d^2}_\theta$ onto an open neighborhood $U$ of $\Uf$ containing the identity matrix. Then the desired result follows from Corollary 2.9, \cite{kirillov2008introduction}. \hfill{$\square$}

Now we have the following proposition.
 
 \begin{proposition}\label{prop:exp}
 Let $\boldsymbol{\mathcal{B}^{(d)}}=\{B^{(d)}_j : 1\leq j\leq d^2\}$ denote a basis of Hermitian unitary matrices for $\C^{d\times d}$ as described in Theorem \ref{thm:basis2} or Corollary \ref{cor:basis}. Then $$\exp(\pm i\theta_jB^{(d)}_j)=\cos \theta_j I_d \pm i\sin \theta_jB^{(d)}_j,$$ for any $\theta_j\in \R,$ $1\leq j\leq d^2.$
 \end{proposition}
 
 \pf The proof follows from the fact that $\exp(\pm it\sigma_j)=\cos t\pm i\sin t\sigma_j,$ $j=0,1,2,3$, $t\in\R,$ and $P_{(k,d-1)}$ is a symmetric unitary matrix, as described in  Theorem \ref{thm:basis2} and Corollary \ref{cor:basis}. \hfill{$\square$}
 
 Thus it follows from Proposition \ref{prop:exp} that exponential of basis elements given in Corollary \ref{cor:basis} is either a $2$-level matrix or a diagonal matrix since the basis elements $U^{2^n}_j,$ $1\leq j\leq 2^n$ are either a $2$-level or a diagonal matrix. It is a well-known result that any unitary matrix can always be written as a product of $2$-level matrices \cite{nielsen2002quantum}.  On the other hand, due to Theorem \ref{thm:span} and Proposition \ref{prop:exp}, it is clear that as a byproduct of the construction of the proposed basis, it provides such a decomposition. 

Note that any unitary matrix is a unit scaling of a special unitary matrix, hence now onward we focus on special unitary matrices. We consider $2$-sparse unitary matrices that are block diagonal matrices with each block is a special unitary matrix. Let $R_a(\theta)$ denote a rotation gate around an axis $a$ with an angle $\theta\in \R.$ In particular, when the rotation matrices around the axes $X, Y, Z$ are defined as $R_Z(\theta)=\bmatrix{e^{i \theta}&0\\0&e^{-i \theta}},$ $R_Y(\theta)=\bmatrix{\cos{\theta} &\sin{\theta}\\-\sin{\theta}&\cos{\theta}},$ $R_X(\theta)=\bmatrix{\cos{\theta} &i \sin{\theta}\\ i \sin{\theta}&\cos{\theta}}.$ 
\begin{definition} \label{def:mcgate}\cite{krol2022efficient}
For $n$-qubit systems, a multi-controlled rotation gate around an axis $a$ is defined as 
\begin{tiny}$$\framebox[7.5cm][c]{\Qcircuit @C=1em @R=.7em {
 &\lstick{1}&\qw& \gate{\circ} &\gate{\circ} &\qw & \hdots\vdots &\gate{\circ} &\gate{\circ}&\qw\\
 &\lstick{2}&\qw&\gate{\circ}\qwx[-1] &\gate{\circ}\qwx[-1] &\qw & \hdots\vdots  &\gate{\circ}\qwx[-1] &\gate{\circ}\qwx[-1] &\qw\\
&\lstick{\vdots}&\qw&\gate{\circ}\qwx[-1] &\gate{\circ}\qwx[-1] &\qw & \hdots\vdots  &\gate{\circ}\qwx[-1] &\gate{\circ}\qwx[-1] &\qw\\
&\lstick{n-1}&\qw& \gate{\circ}\qwx[-1] &\gate{\circ}\qwx[-1] &\qw &\hdots \vdots  &\gate{\circ}\qwx[-1] &\gate{\circ}\qwx[-1] &\qw\\
&\lstick{n}&\qw& \gate{R_a(\theta_1)} \qwx[-1]&\gate{R_a(\theta_2)} \qwx[-1] &\qw& \hdots &\gate{R_a(\theta_{2^{n-2}})} \qwx[-1]&\gate{R_a(\theta_{2^{n-1}})}\qwx[-1] &\qw\\}}$$, \end{tiny}

where \Qcircuit @C=1em @R=.7em {&\gate{\circ}&\qw\\} $\in\{$ \Qcircuit @C=1em @R=.7em {&\ctrl{0}&\qw},\hspace{.5cm}\Qcircuit @C=1em @R=.7em {&\ctrlo{0}&\qw}$\},$ and $\theta_j, 1\leq j\leq 2^{n-1}\in \R.$ Then the unitary matrix corresponding to the above circuit is given by $F_n(R_a(\theta_1,\theta_2,\hdots,\theta_{2^{k-1}}))$, $$F_n(R_a) =\left[ \begin{array}{c|c|c|c} 
      R_a(\theta_1) &  0 &0 & 0\\ 
      \hline 
       0 &  0&\ddots &0\\
        \hline
       0 & 0& 0& R_a(\theta_{2^n-1})
    \end{array} 
    \right] $$
\end{definition} 

Further, it can be shown that the multi-controlled rotation gates can be decomposed and implemented through CNOT and single qubit gates \cite{krol2022efficient}. Indeed, the multi-controlled rotation gate on an $n$ qubit system given by \begin{tiny}\begin{eqnarray}\label{multi1}
\Qcircuit @C=1em @R=.7em {
    &\lstick{1}&\qw& \gate{} &\qw\\
    &\lstick{2}&\qw&\gate{ }\qwx[-1]&\qw\\
    &\lstick{\vdots}&\qw&\gate{ }\qwx[-1]&\qw\\
    &\lstick{n-1}&\qw&\gate{ }\qwx[-1]&\qw\\
    &\lstick{n}&\qw&\gate{F_n(R_a(\psi_1,\hdots,\psi_{2^{n-1}}))}\qwx[-1]&\qw\\}\end{eqnarray}\end{tiny} can be written as
    
    \begin{tiny}\begin{eqnarray}\label{multidec}
   \Qcircuit @C=1em @R=.7em {
    &\lstick{1}&\qw& \qw &\ctrl{4} &\qw& \qw &\ctrl{4} &\qw \\
    &\lstick{2}&\qw&\gate{ }&\qw&\qw&\gate{ }&\qw&\qw\\
    &\lstick{\vdots}&\qw&\gate{ }\qwx[-1]&\qw&\qw&\gate{ }\qwx[-1]&\qw&\qw\\
    &\lstick{n-1}&\qw&\gate{ }\qwx[-1]&\qw&\qw&\gate{ }\qwx[-1]&\qw&\qw\\
    &\lstick{n}&\qw&\gate{F_{n-1}(R_a(\theta_1,\hdots,\theta_{2^{n-1}}))}\qwx[-1]&\targ&\qw&\gate{F_{n-1}(R_a(\phi_1,\hdots,\phi_{2^{n-1}}))}\qwx[-1]&\targ&\qw\\}\end{eqnarray}\end{tiny}  where $a\in\{Y,Z\}$ and $$\psi_j=\begin{cases}
        \theta_j+\phi_j  \mbox{ where } 1\leq j\leq 2^{n-2}\\
        \theta_j-\phi_j \mbox{ where } 2^{n-2}+1\leq j\leq 2^{n-1}.
    \end{cases}$$
\begin{lemma}\label{mzyz}
 The quantum circuits in equation (\ref{multi1}) and Equation (\ref{multidec}) are equivalent when $a\in\{Y,Z\}$.  
\end{lemma}
\pf The proof is computational and easy to verify. \hfill{$\square$}

Now with the help of the multi-controlled rotation gates, we consider writing $2$-sparse block diagonal matrix of the form \begin{eqnarray}\label{blkdg}
\left[ 
\begin{array}{cccc} 
      U_1(\Theta_1) &  & &  \\ 
       &  U_2(\Theta_2) & & \\
       &  & \ddots &   \\
        & & &U_{2^{k-1}}(\Theta_{2^{n-1}})
    \end{array} 
    \right] \end{eqnarray} in terms of the proposed basis elements, where $U_{j}(\Theta_j) \in\Sf\Uf(2),\Theta_j:=(\alpha_j,\beta_j,\gamma_j),$ $1\leq j\leq 2^{n-1}$ is a $2\times 2$ special unitary matrix such that \begin{equation}\label{def:uj}U_{j}(\Theta_j)=\bmatrix{e^{i (\alpha_j+\beta_j)}\cos{\gamma_j}&e^{i (\alpha_j-\beta_j)}\sin{\gamma_j}\\-e^{-i (\alpha_j-\beta_j)}\sin{\gamma_j}&e^{-i (\alpha_j+\beta_j)}\cos{\gamma_j}}.\end{equation}

Since any $2\times 2$ special unitary matrix has a $ZYZ$ decomposition, the matrices in equation (\ref{blkdg}) have circuit from using the multi-controlled rotation gates as \begin{tiny}
\begin{eqnarray}\label{mzyz2}
  \Qcircuit @C=1em @R=.7em {
    &\lstick{1}&\qw& \gate{} &\qw&\qw&\qw& \gate{} &\qw&\qw&\qw& \gate{} &\qw&\qw\\
    &\lstick{\vdots}&\qw& \gate{}\qwx[-1]&\qw &\qw&\qw& \gate{}\qwx[-1]&\qw &\qw&\qw& \gate{}\qwx[-1]&\qw &\qw\\
    &\lstick{n-1}&\qw&\gate{ }\qwx[-1]&\qw&\qw&\qw&\gate{ }\qwx[-1]&\qw&\qw&\qw&\gate{ }\qwx[-1]&\qw&\qw\\
    &\lstick{n}&\qw&\gate{F_n(R_Z)}\qwx[-1]&\qw&\qw&\qw&\gate{F_n(R_Y)}\qwx[-1]&\qw&\qw&\qw&\gate{F_n(R_Z)}\qwx[-1]&\qw&\qw\\}\end{eqnarray} \end{tiny} which we denote as $M_nZYZ$, where $F_n(R_Z)=F_n(R_z(\alpha_1,\hdots,\alpha_{2^{n-1}})),$ $F_n(R_Y)=F_n(R_Y(\gamma_1,\hdots,\gamma_{2^{n-1}})),$ and $F_n(R_Z)=F_n(R_Z(\beta_1,\hdots,\beta_{2^{n-1}}))$  
    
    


\begin{theorem}\label{2mult}
The special unitary matrix corresponding to an $M_nZYZ$ given by equation (\ref{blkdg}) can be written as  $\left(\prod_{l=0}^{2^{n-1}-1} \exp(i t_{l}(\chi_{n-1}^{-1}(l)\otimes \sigma_3)\right)$ $\left(\prod_{j=1}^{2^{n-1}}\exp{(i\theta_{4j^2-2j}U^{(2^n)}_{4j^2-2j})}\right)$ $\left(\prod_{l=0}^{2^{n-1}-1} \exp(i t'_{l}(\chi_{n-1}^{-1}(l)\otimes \sigma_3)\right)$
 where $\theta_{4j^2-2j}=\gamma_{j}\in \R, 1\leq j \leq  2^{n-1}, t_l,t'_l\in \mathbb{R}.$ 
\end{theorem}
\pf The proof is computational and easy to verify. \hfill{$\square$} 

\begin{remark} Note that computing the exponential of Pauli string matrices is a difficult task because the fundamental Pauli matrices do not commute. Besides, in the worst-case scenario, generating a Pauli string for an $n$-qubit system would require $O(n2^{2n})$ operations using generic Kronecker product algorithms \cite{romero2023paulicomposer}. On the contrary, the construction of the proposed basis matrices do not require any operation as the construction is completely prescribed by the pattern of the non-zero entries of the basis elements.    
\end{remark}

\subsection{Approximation of arbitrary unitaries inspired by reduced $CNOT$ gate count}    

We can utilize Theorem \ref{thm:span} to find a value $\theta>0$ such that the set $\left\{\prod_{j=1}^{d^2} \exp\left({i\theta_j B^{(d)}_j}\right)\right\},$ where $(\theta_1,\hdots,\theta_{d^2})\in I^{d^2}_\theta$, generates the unitary group $\Uf(d)$. As a result, any unitary matrix $U$ up to permutation of indices of the basis elements can be represented as 
\begin{small}\begin{eqnarray}\label{eqn:uapprox}  U &=& \underbrace{\left(\prod_{j=1}^{d^2} \exp\left({i\theta_j B^{(d)}_j}\right)\right)\hdots \left(\prod_{j=1}^{d^2} \exp\left({i\theta_j B^{(d)}_j}\right)\right)}_{L\, \mbox{times}} \nonumber \\ &:=& \prod_{l=1}^L \left(\prod_{j=1}^{d^2} \exp\left({i\theta_{lj}B_{j}^{(d)}}\right)\right)\end{eqnarray} \end{small}
for some positive integer $L$, which we call the number of layers or iterations for approximating $U$. However, determining the appropriate value of $L$ for a given $U\in \Uf(d)$ is challenging in practice. 
Indeed, we propose to find a parametric representation of a given unitary through solving the following optimization problem  $$\min_{\theta_{lj}\in I^{Kd^2}_\theta}\left\|U-\prod_{l=1}^L \left(\prod_{j=1}^{d^2} \exp\left({i\theta_{lj}B_{j}^{(d)}}\right)\right)\right\|_F$$ for some $\theta>0,$ where $\|\cdot\|_F$ denotes the Frobenius matrix norm. 

For $n$-qubit system the algorithm faces another significant problem for implementation of the unitary matrices through elementary gates, since the ultimate goal is to implement any unitary through strings of elementary gates. Indeed, for unitary matrices of order $d=2^n,$ the problem with above ordering of multiplication of exponential of basis elements lies in the fact that in order to construct a quantum circuit for the proposed ordering of the basis elements while approximating any unitary from $\Sf\Uf(2^n)$, the number of $CNOT$ gates required for a single iteration becomes $O(2^{3n})$ as follows from equation (\ref{eqn:uapprox}) setting $L=1$. This is due to the fact all non-diagonal RBB matrices generate $2$-level unitary matrices and  a single $2$-level unitary matrix requires at least $2^{n-1}$ $CNOT$ gates from this ordering of the basis elements and there are $2^{2n}-2^n$ non-diagonal basis matrices.

Thus the question is: how to choose a suitable ordering of the basis elements? One motivation for a suitable choice is to reduce the number of $CNOT$ gates in  a quantum circuit implementation of a given unitary matrix using equation (\ref{eqn:uapprox}). First we introduce two functions through which we like to call the proposed basis elements of particular index. We define the functions: $f:\mathbb{N}\times \mathbb{Z}\rightarrow \mathbb{Z}$ and $h:\mathbb{N}\times \mathbb{Z}\rightarrow \mathbb{Z}$ such that
\begin{small}\begin{eqnarray}\label{definition}
\begin{cases}
f(n,k):=f_n(k)=(n-1)^2+(n-1)+(k \mbox{ mod } (n-1))\\
h(n,k):=h_n(k)=(n-1)^2+(k \mbox{ mod } (n-1)).\\
\end{cases}\end{eqnarray} \end{small}


First observe that a SRBB element $U^{(2^n)}_j\in \boldsymbol{\mathcal{U}}^{(2^n)},$ $\exp(i \theta U^{(2^n)}_{j}), j=h_q(p), p<q,q\in \{2,\hdots,2^n\}$, can be written as $\left(\prod_{l=0}^{2^{n-1}-1} \exp(i t_{l}(\chi_{n-1}^{-1}(l)\otimes \sigma_3)\right) \exp{(i \theta U^{(2^n)}_{j'})}$ $\left(\prod_{l=0}^{2^{n-1}-1} \exp(i t'_{l}(\chi_{n-1}^{-1}(l)\otimes \sigma_3)\right)$ where $j'=f_q(p), p<q,q\in \{2,\hdots,2^n\}$ for some $t_l,t'_l\in \mathbb{R}$. Moreover, we would like to consider the ordering of the SRBB such that products of  the exponential certain non-diagonal SRBB elements in that order should generate $M_nZYZ$ type matrix or a block-diagonal special unitary matrix. For example, note from Theorem \ref{2mult} that in the original ordering of the non-diagonal SRBB matrices with indices $4j^2-2j$ and diagonal SRBB matrices, the matrix  $\left(\prod_{l=0}^{2^{n-1}-1} \exp(i t_{l}(\chi_{n-1}^{-1}(l)\otimes \sigma_3)\right)$ $\left(\prod_{j=1}^{2^{n-1}}\theta_{4j^2-2j}U^{(2^n)}_{4j^2-2j}\right)$ $\left(\prod_{l=0}^{2^{n-1}-1} \exp(i t'_{l}(\chi_{n-1}^{-1}(l)\otimes \sigma_3)\right)$ is a $M_nZYZ$ type matrix. Besides, it is well known that quantum circuit for $M_nZYZ$ is prevalent in literature \cite{krol2022efficient}.


From Corollary \ref{cor:basis}, note that any non-diagonal element of the SRBB is given by $U_j^{(2^n)}=PMP,$ where $M$ is a block diagonal matrix with a maximum $3$ blocks, two of which are diagonal matrices and one block is $\sigma\in\{\sigma_1,\sigma_3\},$ and $P=P_{(\alpha,\beta)}$ is a trnasposition with $0<\alpha\leq \beta\leq 2^n.$ 

Observe that for a pair $1\leq \alpha<\beta\leq 2^n$ with $\beta$ is even and $\alpha$ is odd, we have  $P_{(\alpha+1,\beta)}\exp\left(i \theta U_{f_\beta(\alpha)}^{(2^n)}\right)P_{(\alpha+1,\beta)}$ gives a $M_nZYZ$ type matrix. Similarly, if $\beta$ is odd and $\alpha$ is even then $P_{(\alpha,\beta+1)}\exp\left(i \theta U_{f_\beta(\alpha)}^{(2^n)}\right) P_{(\alpha,\beta+1)}$ is a $M_nZYZ$ type matrix. Next if $\alpha$ and $\beta$ are both odd then  $P_{(\alpha+1,\beta)}\exp\left(\iota \theta U_{f_\beta(\alpha)}^{(2^n)}\right)P_{(\alpha+1,\beta)}$ will be a special unitary block-diagonal matrix consists of blocks are of size $2$ belonging to $\Uf(2)$ with at least one block of the form $\bmatrix{\exp(i\theta)&0\\ 0& \exp(i\theta)},$ $n>2.$ Similarly, if $\alpha, \beta$ are even then $P_{(\alpha,\beta-1)}\exp\left(i \theta U_{f_q(p)}^{(2^n)}\right)P_{(\alpha,\beta-1)}$ is a special unitary block diagonal matrix with at least one block is from $\Uf(2).$ Similar observations also hold for the function $h.$ 


Finally observe that all the transpositions $P_{(\alpha,\beta)}$ whose pre and pro multiplication make a matrix $U_j^{(2^n)}\in \boldsymbol{\mathcal{U}}^{(2^n)}$  of type $M_nZYZ$ or a special unitary block diagonal matrix, have the values of $\alpha,\beta$ both to be even or $\alpha$ is even and $\beta$ is odd, where $1\leq \alpha<\beta\leq 2^n.$ 

Thus we consider two sets of permutation matrices \begin{eqnarray*}
\mathcal{P}_{2^n, \mbox{even}} &=& \{P_{(\alpha,\beta)}\in \mathcal{P}_{2^n} \, | \,  \alpha, \beta \,\, \mbox{are even}\} \\
\mathcal{P}_{2^n, \mbox{odd}} &=& \{P_{(\alpha,\beta)} \in \mathcal{P}_{2^n} \,| \, \alpha \, \, \mbox{is even}, \, \beta \,\, \mbox{is odd}\}
\end{eqnarray*} which we will use in order to approximate a unitary matrix as a product of $M_nZYZ$ or unitary block diagonal matrices and its permutations. Then it follows that $\left|\mathcal{P}_{2^n, \mbox{even}}\right|=2^{2n-3}-2^{n-2}=\left|\mathcal{P}_{2^n, \mbox{odd}}\right|.$ Let $T^{e}_x$ and $T^{o}_x$ be sets of $2^{n-2}$ disjoint transpositions from $\mathcal{P}_{2^n, \mbox{even}}$ and $\mathcal{P}_{2^n, \mbox{odd}}$ respectively, $1\leq x\leq 2^{n-1}-1$ such that $\cupdot_x T^e_x=\mathcal{P}_{2^n, \mbox{even}}$ and $\cupdot_x T^o_x=\mathcal{P}_{2^n, \mbox{odd}},$ where $\cupdot$ denotes disjoint union. Define \begin{equation}\label{eqn:pieox}\Pi \mathsf{T}^e_x=\prod_{(\alpha,\beta)\in T^e_x} P_{(\alpha,\beta)} \,\, \mbox{and} \,\, \Pi \mathsf{T}^o_x=\prod_{(\alpha,\beta)\in T^o_x} P_{(\alpha,\beta)},\end{equation} $1\leq x\leq 2^{n-1}-1.$

The motivation behind creating unitary block diagonal or $M_nZYZ$ matrices lies in the fact that the quantum circuits for such matrices are easy to implement. The quantum circuit for $M_nZYZ$ matrices can be found in \cite{krol2022efficient}. We shall see later that adding a few $CNOT$ and $R_Z$ gates it is possible to define a circuit for a special unitary block diagonal matrix with $2\times 2$ blocks from a circuit that represents a $M_nZYZ$ matrix. Then we have the following result.


\begin{lemma}\label{blockdiagonal2}
    A block diagonal matrix $U\in \Sf\Uf(2^n)$ consisting of $2\times 2$ blocks is of the form $\left(\prod_{t=2}^{2^n} \exp\left(i \theta_{t^2-1} U^{(2^n)}_{t^2-1}\right)\right)$ $\left(\prod_{j=1}^{2^{n-1}}\exp\left(i\theta_{4j^2-2j} U^{(2^n)}_{4j^2-2j}\right)\right)$ $\left(\prod_{t=2}^{2^n} \exp\left(i \theta_{t^2-1} U^{(2^n)}_{t^2-1}\right)\right)$ where $\theta_{4j^2-2j}\in \mathbb{R},1\leq j\leq 2^{n-1}, \theta_{t^2-1},\theta'_{t^2-1}$ are obtained from Theorem\ref{2mult}.
\end{lemma}
\pf The proof is computational and easy to verify. \hfill{$\square$}

Now from equation (\ref{eqn:pieox}), for any $1\leq x\leq 2^{n-1}-1,$ define 
\begin{small}\begin{eqnarray}
M^e_x &=& \Pi \mathsf{T}^e_x  \left[  \prod_{(\alpha,\beta)\in T^e_x} \exp\left(i \theta_{h_{\beta-1}(\alpha)} U^{(2^n)}_{h_{\beta-1}(\alpha)}\right) \right. \nonumber \\ 
&& \exp\left(i \theta_{f_{\beta-1}(\alpha)} U^{(2^n)}_{f_{\beta-1}(\alpha)}\right)  \exp\left(i \theta_{h_{\beta}(\alpha-1)} U^{(2^n)}_{h_{\beta}(\alpha-1)}\right)    \nonumber \\ 
&& \left. \exp\left(i \theta_{f_{\beta}(\alpha-1)} U^{(2^n)}_{f_{\beta}(\alpha-1)}\right) \right] \Pi \mathsf{T}^e_x, \label{def:mex} \\
M^o_x &=& \Pi \mathsf{T}^o_x  \left[  \prod_{(\alpha,\beta)\in T^o_x} \exp\left(i \theta_{h_{\beta}(\alpha-1)} U^{(2^n)}_{h_{\beta}(\alpha-1)}\right)  \right.\nonumber \\ 
&& \exp\left(i \theta_{f_{\beta}(\alpha-1)} U^{(2^n)}_{f_{\beta}(\alpha-1)}\right) \exp\left(i \theta_{h_{\beta+1}(\alpha)} U^{(2^n)}_{h_{\beta+1}(\alpha)}\right)  \nonumber \\ 
&& \left. \exp\left(i \theta_{f_{\beta+1}(\alpha)} U^{(2^n)}_{f_{\beta+1}(\alpha)}\right) \right] \Pi \mathsf{T}^o_x, \label{def:mox}
\end{eqnarray}\end{small}
where $\Pi \mathsf{T}^e_x$ and $\Pi \mathsf{T}^o_x$ are defined in equation (\ref{eqn:pieox}). Then it can be seen that $M^o_x\in \Sf\Uf(2^n)$ is a  block-diagonal matrix with $2\times 2$ blocks and $M^e_x\in \Sf\Uf(2^n)$ is a $M_nZYZ$ matrix, $1\leq x\leq 2^{n-1}-1.$

{\bf Approximation of unitary matrices with optimal ordering:} Define \begin{eqnarray}
\zeta(\Theta_{\zeta}^{(l)}) &=& \prod_{j=2}^{2^n} \exp\left(i \theta_{j^2-1}^{(l)} U^{(2^n)}_{j^2-1}\right) \label{eqn:thetazeta} \\
\Psi(\Theta_{\psi}^{(l)}) &=&  \left(\prod_{j=1}^{2^{n-1}} \exp\left(i \theta_{(2j-1)^2}^{(l)} U^{(2^n)}_{(2j-1)^2}\right) \right) \nonumber \\ 
&& \exp\left(i \theta_{(4j^2-2j)}^{(l)} U^{(2^n)}_{(4j^2-2j)}\right) \nonumber \\ 
&& \left(\prod_{x=1}^{2^{n-1}-1}  \left(\Pi \mathsf{T}^e_x\right) M^e_x \left(\Pi \mathsf{T}^e_x\right)\right)\label{eqn:thetapsi}\\
\Phi(\Theta_{\phi}^{(l)}) &=& \left(\prod_{x=1}^{2^{n-1}-1} \left(\Pi \mathsf{T}^o_x\right) M^o_x \left(\Pi \mathsf{T}^o_x\right)\right). \label{eqn:thetaphi}
\end{eqnarray} Then note that $\zeta(\Theta_{\zeta}^{(l)})$ is the product of exponential of all diagonal SRBB elements. Besides, it is computational to check that $\Psi(\Theta_{\psi})$ is the product of matrices of type $M_nZYZ$ and permutation scaling of $M_nZYZ$ type matrices, and $\Phi(\Theta_{\phi}^{(l)})\in \Sf\Uf(2^n)$ is product of  block-diagonal matrices, which we will use in the construction of the circuits for these matrices in Section \ref{sec:circuit}. Then we propose a quantum neural network framework \cite{benedetti2019parameterized} for approximating a unitary matrix as follows. Given $U\in\Sf\Uf(2^n)$, approximate $U$ as \begin{equation}\label{def:uapprox}
    U \equiv \prod_{l=1}^L \zeta(\Theta_{\zeta}^{(l)})\, \Psi(\Theta_{\psi}^{(l)})\, \Phi(\Theta_{\phi}^{(l)}) 
\end{equation}
where $l$ is called the layer and we call the equation (\ref{def:uapprox}) the $L$-layer approximation of $U.$



\begin{algorithm}
\caption{Algorithm for Approximating $2^n\times 2^n$ special unitary matrix}
\textbf{Provided:} $U_1\in \Sf\Uf(2^n)$, $U^{(2^n)}_j \in \boldsymbol{\mathcal{U}}^{(2^n)}, 1\leq j\leq 2^{2n}-1,$ $\zeta(\Theta_{\zeta}),$ $\Psi(\Theta_{\psi}),$ $\Phi(\Theta_{\phi})$ given by equation (\ref{eqn:thetazeta}) - (\ref{eqn:thetaphi}).

\textbf{Input:} $\Theta_{\zeta}^{(0)}$, $\Theta_{\psi}^{(0)},$ $\Theta_{\phi}^{(0)},$ $\epsilon>0$ 
 \\
\textbf{Output:} $A=\prod_t \zeta(\Theta^{(t)}_{\zeta}) \Psi(\Theta^{(t)}_{\psi}) \Phi(\Theta^{(t)}_{\phi})$ such that $\|U-A\|_F\leq \epsilon$
\begin{algorithmic}
\Procedure{}{Unitary Matrix $U$}
\State $A\rightarrow I$
\For{$t=1;t++$}
\State Use an optimization method like Nelder-Mead/Powell's or Gradient descent method to find $\Theta_{\zeta}^{(t)}$, $\Theta_{\psi}^{(t)},$ $\Theta_{\phi}^{(t)}$ such that $$\min_{\Theta_{\zeta}^{(t)},\Theta_{\psi}^{(t)},\Theta_{\phi}^{(t)}} \left\|U-\zeta(\Theta^{(t)}_{\zeta}) \Psi(\Theta^{(t)}_{\psi}) \Phi(\Theta^{(t)}_{\phi})\right\|_F=\epsilon_t$$
\If{$\epsilon_t\leq \epsilon$}
\State \textbf{Break}
\State $A \rightarrow A\zeta(\Theta^{(t)}_{\zeta}) \Psi(\Theta^{(t)}_{\psi}) \Phi(\Theta^{(t)}_{\phi})$
\Else
\State $U_{t+1}\rightarrow U_{t}A^*$\\
\EndIf
\State \textbf{End}
\EndFor
\State \textbf{End}
\State \textbf{End Procedure}
\EndProcedure
\end{algorithmic}
\end{algorithm}

\subsection{Numerical simulations}

In this section we report the performance of the proposed algorithms for approximating unitary matrices through product of exponential of the proposed RBB elements in optimal ordering. Given a target unitary matrix, the initial choice of the parameters can influence the output unitary matrix and since the objective function is non-convex, the optimal approximated values of the parameters may lead to a local minimum. Thus we generate up to $10^3$ random points from uniform distribution and normal distribution for the set of parameters $\Theta=\{\theta_1,\hdots,\theta_{2^{2n}-1}\},$ where $0 \leq \theta_j\leq 2\pi, 1\leq j\leq 2^{2n}-1$ and execute the proposed algorithms. Finally, we report the error that is least among all those initial parameter values. 

\begin{figure}[ht]
  \centering
  \includegraphics[width=0.8\linewidth]{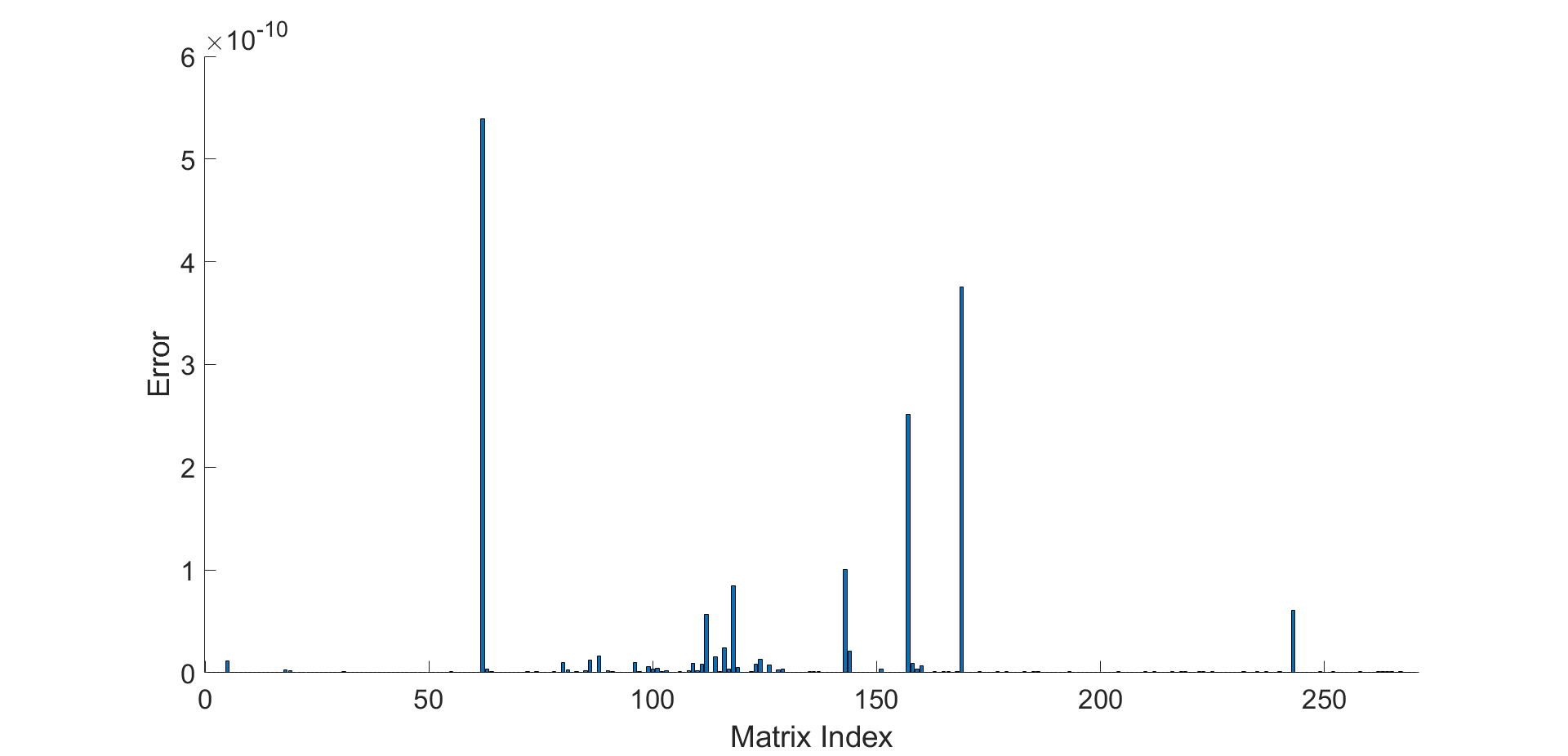}  
\caption{Approximation errors using Algorithm 1 for $2$-qubit unitary matrices sampled from random Haar distribution.}\label{fig:error2qubit}
\label{fig:randerror}
\end{figure}

\begin{figure}[ht]
  \centering
  \includegraphics[width=0.7\linewidth]{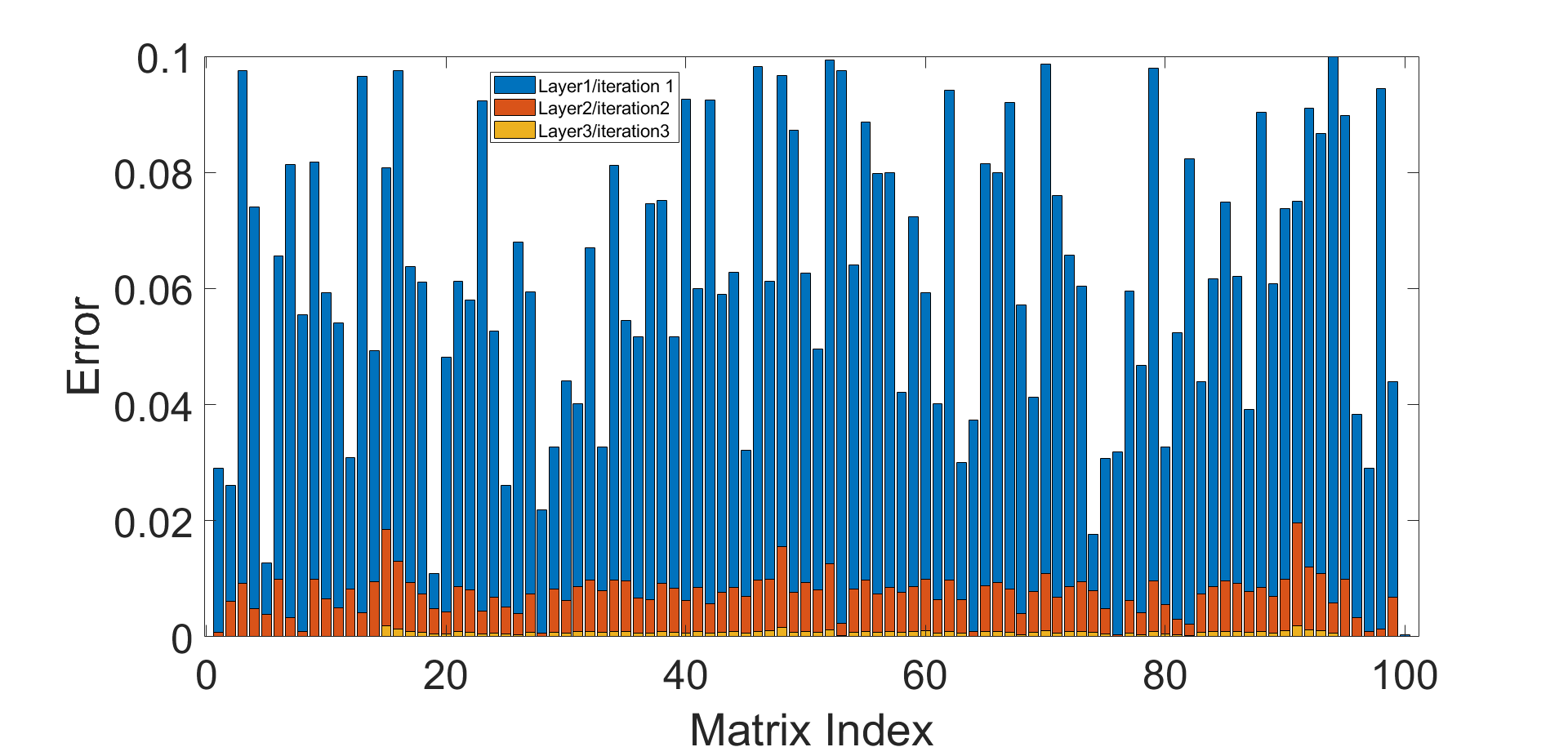}  
  \caption{The errors obtained from up to three iterations (layers) for $3$-qubit Haar random unitaries. The error after $3$rd iteration lies between $10^{-4}$ to $10^{-6}.$}
  \label{fig:error3qubit}
\end{figure}

\begin{figure}[ht]
  \centering
  \includegraphics[width=0.7\linewidth]{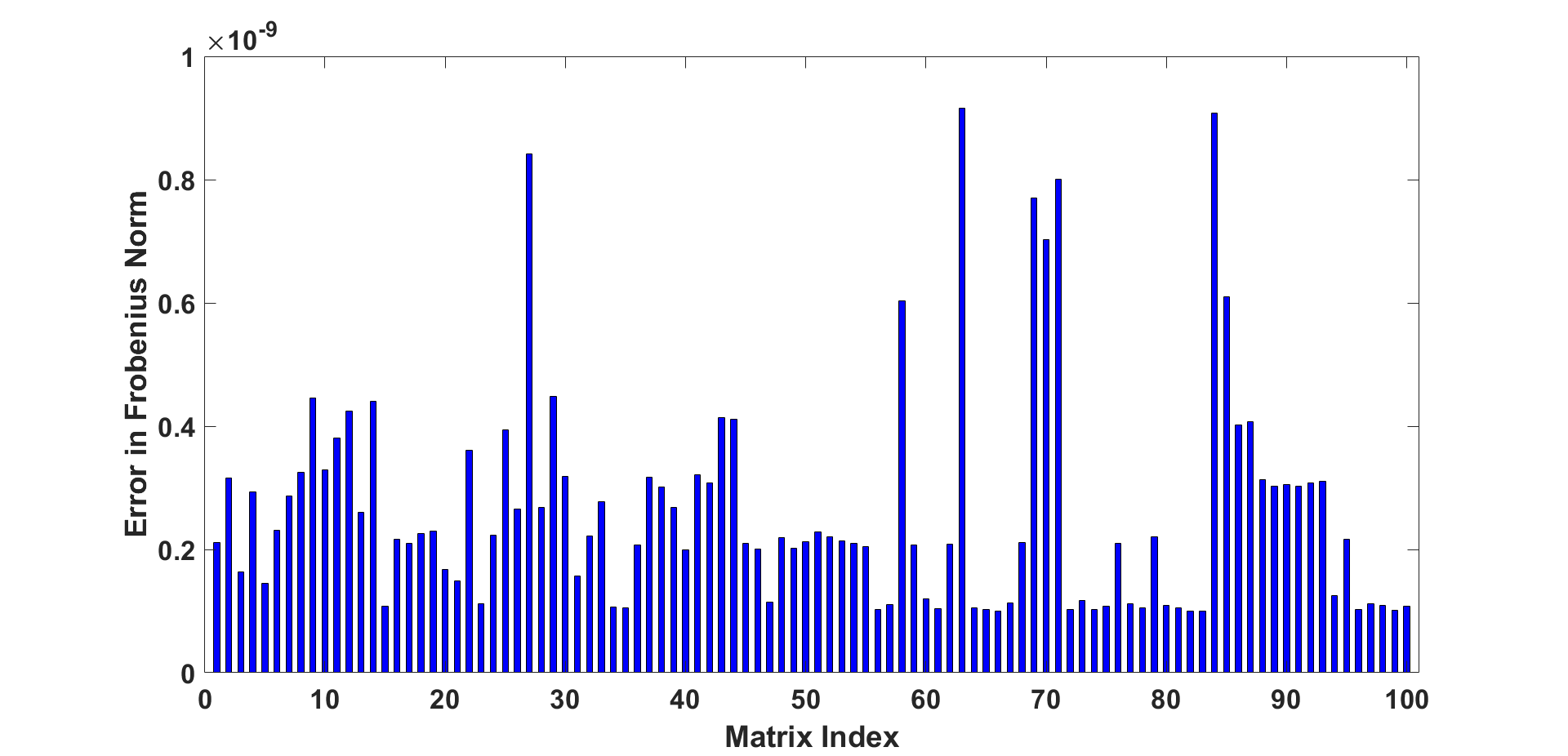}  
\caption{Errors for simulating random $8$-sparse $4$-qubit block-diagonal unitaries considering only one iteration of the Algorithm 1.}
\label{fig:error3qubit2}
\end{figure}

We compare our findings with the results found in \cite{krol2022efficient,vidal2004universal,kraus2001optimal,younis2020qfast} and see that our method for $2$-qubits is faster as it does not need to perform singular value decomposition. Like \cite{vidal2004universal,kraus2001optimal}, we don't need to convert the target matrices into magic basis/states and perform Schmidt decomposition in order to check for separable states which is non-trivial and time consuming. We have also seen that employing the modified ordering of the proposed basis elements and decomposing a $2$-qubit gate using the original ordering of the basis elements with $(\prod_{j=1}^{15}\exp{(i \theta_i U^{(4)}_i)})^L$ with $L=1$ and applying Algorithm 1, the error is same. We have performed Algorithm 1 on MATLAB and Python 3.0 on a system with $16GB$ RAM, Intel(R) Core(TM) $i5-1035G1$ CPU $@ 1.00GHz 1.19 GHz$ for $2$-qubit and $3$-qubit examples. For $4$-qubit examples, we have performed the simulations using supercomputer PARAM Shakti of IIT Kharagpur.  The approximation errors are reported in Tables \ref{Table:error2q}, \ref{table:error3q}, \ref{table:error4q} and Figures \ref{fig:randerror}, \ref{fig:error3qubit}, \ref{fig:error3qubit2}. 



\begin{table}
\begin{center}
\begin{tabular}{ |c|c|c|c| }
\hline
 Matrix &Time taken  & Error from & Error from \cite{vidal2004universal}   \\
 &in seconds & our Method & circuit + \\
 &in our method&& our method\\
\hline \hline
CNOT&90 &$7.977 \times 10^{-14}$ & $1.2\times10^{-15}$\\
\hline
$Grover_2$&  124 & $1.256\times 10^{-15}$  & $3.66\times10^{-13}$\\
\hline
XX&  20& $6.226\times 10^{-12}$& $4.2\times10^{-13}$\\
\hline
YY&  240&$3.223\times 10^{-15}$ & $1.5\times10^{-13}$\\
\hline
ZZ  & 90&$1.363\times 10^{-17}$& $2.96\times10^{-13}$\\
\hline
SWAP  & 63&$1.839\times 10^{-13}$ & $6.8\times10^{-16}$\\
\hline
XZ & 150& $3.580\times 10^{-13}$ & $9.884\times10^{-13}$\\
\hline
ZX & 129& $5.438\times 10^{-13}$ &$4.332\times10^{-13}$\\
\hline
ZY& 121 & $3.188\times 10^{-12}$ & $1.065\times10^{-13}$\\
\hline
CNOT(2,1)& 45& $1.058\times 10^{-13}$ & $2.89\times10^{-13}$\\
\hline
DCNOT & 29& $4.020\times 10^{-13}$ & $4.92\times10^{-14}$\\
\hline
XNOR & 23& $3.166\times 10^{-13}$ & $8.04\times10^{-12}$\\
\hline
iSWAP& 183 & $3.003\times 10^{-14}$ &$4.13\times10^{-13}$\\
\hline
fSWAP & 93& $2.037\times 10^{-13}$ & $3.32\times10^{-13}$\\
\hline
C-Phase & 15& $7.666\times 10^{-15}$ & $5.06\times10^{-13}$\\
\hline
XX+YY & 124& $1.665\times 10^{-12}$ & $4.186\times10^{-13}$\\
\hline
$\sqrt{SWAP}$ & 97& $1.686\times 10^{-13}$ & $4.05\times10^{-13}$\\
\hline
$\sqrt{iSWAP}$ &10 & $1.106\times 10^{-13}$ & $2.76\times10^{-13}$\\
\hline
$QFT_2$& 31 & $3.215\times 10^{-13}$ & $9.526\times10^{-13}$\\
\hline
\end{tabular}
\caption{Error and time for simulating standard $2$-qubit unitaries}
\label{Table:error2q}
 \end{center}
\end{table}

\begin{table}
\begin{center}
\begin{tabular}{ |l|l|l|l|l|l|l| }
\hline
Matrix & 1st iteration   & QFAST  & UniversalQ   \\
&Error from  & + KAK  & \cite{younis2020qfast} \\
& our method & \cite{younis2020qfast}&\\
\hline \hline
Toffoli&$4.48 \times 10^{-9}$ & $1.5\times 10^{-6}$ & $2.6\times 10^{-8}$\\
\hline
Fredkin& $1.6 \times 10^{-8}$ & $2.2\times 10^{-6}$&$0$\\
\hline
$Grover_3$& $4.60 \times 10^{-9}$ & $8.1\times 10^{-7}$&$0$ \\
\hline
Peres&$2 \times 10^{-8}$ & $6.8\times 10^{-7}$&$2.1\times 10^{-8}$\\
\hline
$QFT_3$&$ 3.1 \times 10^{-9}$& $3\times 10^{-7}$&$3\times 10^{-8}$ \\
\hline
\end{tabular}
\caption{Error in the Frobenius norm after simulation using one iteration/Layer for $3$-qubit standard unitaries}
\label{table:error3q}
\end{center} 
\end{table}

\begin{table}
\begin{center}
\begin{tabular}{|c|c|c|c|c|c|c|}
\hline
Matrix & 1st iteration  & QFAST + & QFAST +  \\
& Error  from  & KAK \cite{younis2020qfast} & UQ \cite{younis2020qfast}  \\
& our method && \\
\hline \hline
CCCX &$1.97 \times 10^{-8}$ &  $2.2\times 10^{-5}$ & $1.3\times 10^{-6}$ \\
\hline
$Grover_4$& $2.12 \times 10^{-9}$ &  $-$ & $-$ \\
\hline
$QFT_4$&$ 9.331 \times 10^{-8}$ &  $7.9\times 10^{-7}$ & $8.5\times 10^{-7}$ \\
\hline
\end{tabular}
\caption{Error in the Frobenius norm after simulation using one iteration/Layer for $4$-qubit standard unitaries}
\label{table:error4q}
 \end{center} 
\end{table}

\section{Quantum circuit representation of unitary matrices}\label{sec:circuit}

In the previous section we have introduced a modified ordering while multiplying for approximation of unitary matrices of $n$-qubit systems. The modified ordering is introduced to incorporate a structure for approximating a target unitary through product of permutation matrices, $M_nZYZ$ type matrices, and special unitary block-diagonal matrices when we write a given target unitary as product of exponential of SRBB elements through a neural network framework.

\subsection{Quantum circuits for product of transpositions}

Now we construct quantum circuit for the matrix $\Pi\mathsf{T}^s_x,$ $s\in\{e,o\},$ $1\leq x\leq 2^{n-1}-1.$  First consider an $n$-qubit quantum circuit consisting of only $(CNOT)_{(n,i)},$ $1\leq i\leq n-1$ gates as follows. Let $x\equiv (x_1x_2x_3,\hdots x_{n-1})$ denote the binary representation of $0\leq x\leq 2^{n-1}$ i.e. $x=\sum_{i=1}^{n-1}2^{n-i-1}x_i$ where $x_i\in \{0,1\}$. Then for a given $x$ with its binary representation define a circuit with the property that for each $1\leq i\leq n-1$ the circuit contains a $(CNOT)_{(n,i)}$ gate if the $x_i=1,$ where $(CNOT)_{(n,i)}$ means a $CNOT$ gate with $n$-th qubit as the control and $i$-th qubit as target. Note that binary strings $(x_1,x_2,\hdots,x_{n-1})$ and all subsets of $[n-1]=\{1,\hdots,n-1\}$ have a one-one correspondence.  We denote this function as $\chi: \{0,1\}^{n-1}\rightarrow 2^{[n-1]},$ which assigns $x=(x_1,x_2,\hdots,x_{n-1})$ to $\chi(x)=\Lambda=\{j: x_j=1, 1\leq j\leq n-1\}.$ Thus each position of the string represents a characteristic function for $\Lambda.$

\begin{theorem}\label{constructeven}
Let $\chi: \{0,1\}^{n-1}\rightarrow 2^{[n-1]}$ be the bijective function as defined above. Then 
$\Pi\mathsf{T}^e_x=\prod_{m=0}^{2^{n-2}-1} P_{(\alpha_{\Lambda}(m),\beta_{\Lambda}(m))},$ where $m\equiv (m_1,m_2,\hdots,m_{n-1}),$ $\alpha_{\Lambda} (m)= \sum_{i\in \Lambda}m_i2^{n-i}+\sum_{j\not\in \Lambda }m_j2^{n-j}+2,$ $\beta_{\Lambda}(m) = \sum_{i\in \Lambda} \overline{m}_i2^{n-i}+\sum_{j\not\in \Lambda} m_j2^{n-j}+2,$ and $\Pi\mathsf{T}^e_x\neq \Pi\mathsf{T}^e_y$ if $x\neq y.$ 
\end{theorem}
\pf The proof follows from the $n$-qubit representation given by the circuit in equation (\ref{cr:pie})
 \begin{tiny}\begin{equation}\label{cr:pie}\Qcircuit @C=1em @R=.7em {
    &\lstick{1}&\qw&\qw&\qw &\qw&\qw&\qw\\
    &\lstick{\vdots}&\qw&\qw&\qw &\qw&\qw&\qw\\
    &\lstick{m_1}&\targ&\qw&\qw &\qw&\qw&\qw\\
    &\lstick{\vdots}&\qw&\qw&\qw &\qw&\qw&\qw\\
    &\lstick{m_2}&\qw&\targ&\qw &\qw&\qw&\qw\\
    &\lstick{\vdots}&\qw&\qw&\qw &\qw&\qw&\qw\\
    &\lstick{m_3}&\qw&\qw&\targ &\qw&\qw&\qw\\
    &\lstick{\vdots}&\qw&\qw&\qw &\qw&\qw&\qw\\
    &\lstick{\vdots}&\qw&\qw&\qw &\qw&\qw&\qw\\
    &\lstick{m_k}&\qw&\qw&\qw &\targ&\qw&\qw\\
    &\lstick{\vdots}&\qw&\qw&\qw &\qw&\qw&\qw\\
    &\lstick{n}&\ctrl{-9}&\ctrl{-7}&\ctrl{-5} &\ctrl{-2}&\qw&\qw\\}\end{equation}\end{tiny}

\begin{theorem}\label{constructodd}
Let $\chi: \{0,1\}^{n-1}\rightarrow 2^{[n-1]}$ be the bijective function as defined above. Then 
$\Pi\mathsf{T}^o_x=\prod_{m=0}^{2^{n-2}-1} P_{(\alpha_{\Lambda}(m),\beta_{\Lambda}(m))},$ where $m\equiv (m_1,m_2,\hdots,m_{n-1}),$ $\alpha_{\Lambda} (m)= \sum_{i\in \Lambda}m_i2^{n-i}+\sum_{j\not\in \Lambda }m_j2^{n-j}+2,$ $\beta_{\Lambda}(m) = \sum_{i\in \Lambda} \overline{m}_i2^{n-i}+\sum_{j\not\in \Lambda} m_j2^{n-j}+1,$ and $\Pi\mathsf{T}^o_x\neq \Pi\mathsf{T}^o_y$ if $x\neq y.$ 
\end{theorem}
\pf The proof follows from the $n$-qubit representation given by the circuit in equation (\ref{cr:pio})
\begin{tiny}\begin{equation}\label{cr:pio}
    \Qcircuit @C=1em @R=.7em {
    &\lstick{1}&\qw&\qw&\qw&\qw &\qw&\qw&\qw\\
     &\lstick{\vdots}&\qw&\qw&\qw&\qw &\qw&\qw&\qw\\
    &\lstick{m_1}&\ctrl{8}&\targ&\qw&\qw &\qw&\qw&\ctrl{8}\\
    &\lstick{\vdots}&\qw&\qw&\qw&\qw &\qw&\qw&\qw\\
    &\lstick{m_2}&\qw&\qw&\targ&\qw &\qw&\qw&\qw\\
    &\lstick{\vdots}&\qw&\qw&\qw&\qw &\qw&\qw&\qw\\
    &\lstick{m_3}&\qw&\qw&\qw&\targ &\qw&\qw&\qw\\
    &\lstick{\vdots}&\qw&\qw&\qw&\qw &\qw&\qw&\qw\\
    &\lstick{m_k}&\qw&\qw&\qw&\qw &\targ&\qw&\qw\\
    &\lstick{\vdots}&\qw&\qw&\qw&\qw &\qw&\qw&\qw\\
    &\lstick{n}&\targ&\ctrl{-8}&\ctrl{-6}&\ctrl{-4} &\ctrl{-2}&\qw&\targ\\}
\end{equation}\end{tiny}

\subsection{Quantum circuit for diagonal unitaries}

The SRBB basis elements that are diagonal matrices, are given by $U_{j^2-1}^{(2^n)}, 2\leq j\leq 2^n$, which are of the form $\otimes_{j=1}^n A_j, A_j\in\{I_2, \sigma_3\}.$ Given such a basis elment for some $j$, let $m$ be the greatest number such that $A_p=I_2$ for all $p>m$, and let $A_{m_1}=A_{m_2}=\hdots,A_{m_k} =\sigma_3$ for some $k$ with $m_1< m_2 < \hdots <m_k < m.$ Then a quantum circuit representation of $\exp\left(i \theta U^{(2^n)}_{j^2-1}\right)$ is given by  

\begin{tiny}\begin{equation}\label{diag}
{\Qcircuit @C=1em @R=.7em {
    &\lstick{1}& \qw &\qw &\qw &\qw &\qw &\qw &\qw &\qw\\
    &\lstick{\vdots}& \qw &\qw &\qw &\qw &\qw &\qw &\qw &\qw\\
    &\lstick{m_1}& \qw &\ctrl{6} &\qw &\qw &\qw &\qw &\ctrl{6} &\qw\\
    &\lstick{\vdots}& \qw &\qw &\qw &\qw &\qw &\qw &\qw &\qw\\
    &\lstick{m_2}& \qw &\qw &\ctrl{4} &\qw &\qw &\ctrl{4} &\qw &\qw\\
    &\lstick{\vdots}& \qw &\qw &\qw&\qw &\qw &\qw &\qw &\qw\\
    &\lstick{m_k}& \ctrl{2} &\qw &\qw &\qw &\qw &\qw &\qw &\ctrl{2}\\
    &\lstick{\vdots}& \qw &\qw &\qw&\qw &\qw &\qw &\qw &\qw\\
    &\lstick{m}& \targ &\targ &\targ &\gate{R_Z(\theta)} &\qw &\targ &\targ &\targ\\
    &\lstick{\vdots}& \qw &\qw &\qw&\qw &\qw &\qw &\qw &\qw\\
    &\lstick{n}& \qw &\qw &\qw&\qw &\qw &\qw &\qw &\qw\\}}\end{equation} \end{tiny}

\subsection{Quantum circuit for unitary block diagonal matrices}

\begin{corollary}\label{constructblkdg}
   A quantum circuit for a block diagonal matrix $U\in \Sf\Uf(2^n)$ of the form $$\left[ 
\begin{array}{c|c|c|c|c} 
      {U}_2 &  0 &0 & 0&0\\ 
      \hline 
      0&U_4 &0&0&0 \\
      \hline
       0 & 0& 0&\ddots &0\\
        \hline
       0 & 0& 0&0& {U}_{2^{n}}
    \end{array} 
    \right],$$ where $U_{2j}\in \Uf(2), 1\leq j\leq 2^{n-1}$, requires at most $5.2^{n-1}-6$ $CNOT$ gates. 
\end{corollary}
\pf The proof follows from the quantum circuit given by equation (23).   


\subsection{Scalable quantum circuits for approximating special unitary matrices}

 From Algorithm 1, we see that we a special unitary matrix $U\in \Sf\Uf(2^n)$ can be approximated in the circuit form with one layer is given by 
\begin{tiny}\begin{equation}\label{alg3}
\Qcircuit @C=1em @R=.7em {
&\lstick{1}& \multigate{3}{\Phi(\Theta_{\phi})}&\multigate{3}{\Psi(\Theta_{\psi})}&\multigate{3}{\zeta(\Theta_{\zeta})}&\qw\\
&\lstick{\vdots}& \ghost{{\phi(\Theta)}}&\ghost{{\psi(\Theta)}}&\ghost{{\zeta(\Theta)}}&\qw\\
&\lstick{n-1}& \ghost{{\phi(\Theta)}}&\ghost{{\psi(\Theta)}}&\ghost{{\zeta(\Theta)}}&\qw\\
&\lstick{n}& \ghost{{\phi(\Theta)}}&\ghost{{\psi(\Theta)}}&\ghost{{\zeta(\Theta)}}&\qw\\}
\end{equation}\end{tiny}

for $1\leq x\leq 2^{n-1}-1$, $\Pi\mathsf{T}^s_{x} M^s_x \Pi\mathsf{T}^s_x,$ $s\in\{e,o\}$ can have the quantum circuit representation as

    \begin{tiny}
$$\Qcircuit @C=1em @R=.7em {
    &\lstick{1}&\multigate{3}{\Pi\mathsf{T}^s_{x}}& \multigate{3}{M^s_x}&\multigate{3}{\Pi\mathsf{T}^s_{x}}&\qw\\
    &\lstick{\vdots}&\ghost{P_{(x,2^n,odd)}}& \ghost{D'_x}&\ghost{P_{(x,2^n,odd)}} &\qw\\
    &\lstick{\vdots}&\ghost{P_{(x,2^n,odd)}}& \ghost{D'_x}&\ghost{P_{(x,2^n,odd)}} &\qw\\
    &\lstick{n}&\ghost{P_{(x,2^n,odd)}}& \ghost{D'_x}&\ghost{P_{(x,2^n,odd)}} &\qw\\}$$ \end{tiny} where the circuit representation of $M^s_x$ is  of the form in equation (\ref{multi1}), and the circuits for $\Pi\mathsf{T}^s_{x}$ can be determined by Theorem \ref{constructodd} and Theorem \ref{constructeven}.


Now we present the following algorithms based on the above discussion  that will help us to create an algorithm for constructing scalable quantum circuits.

\begin{algorithm}
\caption{Creating circuit for $(n+1)$-qubit rotation gates $F_{(n+1)}(R_z)$ from multi-qubit rotation gates $F_{(n)}(R_z)$}\label{multialgoz}
\textbf{Provided:} $CNOT$ gates,circuits $F_n(R_z)$ .\\
\textbf{Input:}$a_1,a_2,a_4\hdots,a_{2^{n-1}}$ for $Fn(R_z):=F_n(R_z(a_1,a_2\hdots,a_{2^{n-1}}))$ and $b_1,b_2,b_3\hdots,b_{2^{n-1}}$ for $Fn(R_z):=F_n(R_z(b_1,b_2\hdots,b_{2^{n-1}}))$ \\
\textbf{Output:}$\xi(F_n(R_z(a_1,\hdots,a_{2^{n-1}})),F_n(R_z(b_1,\hdots,b_{2^{n-1}}))):=\xi(F_n(R_z),F_n(R_z))$ gives a circuit of $F_{n+1}(R_z)$
\begin{algorithmic}
\State Add one layer of qubit at the top. Add a $(CNOT)_{(1,n+1)}$ to the left of $I_2\otimes F_n(R_z)$. Then add another $(CNOT)_{(1,n+1)}$ and a $I_2\otimes F_n(R_z)$. 
\State End
\end{algorithmic}
\end{algorithm}

\begin{algorithm}
\caption{Creating circuit for $(n+1)$-qubit rotation gates $F_{(n+1)}(R_y)$ from multi-qubit rotation gates $F_{(n)}(R_y)$}\label{multialgoy}
\textbf{Provided:} $CNOT$ gates,circuits $F_n(R_y)$ .\\
\textbf{Input:}$a_1,a_2,a_4\hdots,a_{2^{n-1}}$ for $Fn(R_y):=F_n(R_y(a_1,a_2\hdots,a_{2^{n-1}}))$ and $b_1,b_2,b_3\hdots,b_{2^{n-1}}$ for $Fn(R_y):=F_n(R_y(b_1,b_2\hdots,b_{2^{n-1}}))$ \\
\textbf{Output:}$\xi(F_n(R_y(a_1,\hdots,a_{2^{n-1}})),F_n(R_y(b_1,\hdots,b_{2^{n-1}}))):=\xi(F_n(R_y),F_n(R_y))$ gives a circuit of $F_{n+1}(R_z)$
\begin{algorithmic}
\State Add one layer of qubit at the top. Add a $(CNOT)_{(1,n+1)}$ to the left of $I_2\otimes F_n(R_y)$. Then add another $(CNOT)_{(1,n+1)}$ and a $I_2\otimes F_n(R_y)$. 
\State End
\end{algorithmic}
\end{algorithm}

\begin{algorithm}
\caption{Creating circuit for $\Pi\mathsf{T}^e_{n+1,y}, 0\leq y\leq 2^n-1$ from circuit $\Pi\mathsf{T}^e_{n,x},0\leq x\leq 2^{n-1}-1$}\label{Createeven}
\textbf{Provided:} $CNOT$ gates, circuits $\Pi\mathsf{T}^e_{n,x},0\leq x\leq 2^{n-1}-1$ .\\
\textbf{Input:} $y\in \{0,\hdots,2^n-1\}$\\
\textbf{Output:} $\eta(y,2^{n+1},even)$ gives a circuit of $I_2\otimes \Pi\mathsf{T}^e_{n+1,y}$
\begin{algorithmic}
\For {$y=0:2^{n}-1;y++$}
\If {$y<2^{n-1}$}
\State $x=y$
\State $\eta(y,2^{n+1},even)$ $\rightarrow$  Add one qubit layer at the top.  
\Else
\State $x=y-2^{n-1}$
\State $\eta(y,2^{n+1},even)\rightarrow$ Add one qubit layer at the top and add a $(CNOT)_{(n+1,1)}$ to left of $\Pi\mathsf{T}^e_{n,x}$. 
\EndIf
\EndFor
\end{algorithmic}
\end{algorithm}

\begin{algorithm}
\caption{Creating circuit for $\Pi\mathsf{T}^o_{n+1,y},0\leq y\leq 2^n-1$ from circuit $\Pi\mathsf{T}^o_{n,x},0\leq x\leq 2^{n-1}-1$}\label{Createodd}
\textbf{Provided:} $CNOT$ gates,circuits $\Pi\mathsf{T}^e_{n,x},\Pi\mathsf{T}^o_{n,x},0\leq x\leq 2^{n-1}-1$ .\\
\textbf{Input:} $y\in \{0,\hdots,2^n-1\}$\\
\textbf{Output:}$\eta(y,2^{n+1},odd)$ gives a circuit of $\Pi\mathsf{T}^o_{n+1,x}$
\begin{algorithmic}
\For{$y=0:2^{n}-1;y++$}
\If{$y<2^{n-1}$}
\State $x=y$
\State $\eta(y,2^{n+1},odd)\rightarrow$  Add one qubit layer at the top. 
\Else
\State $x=y-2^{n-1}$
\State $\eta(y,2^{n+1},odd)\rightarrow$ Add one qubit layer at the top and add a $(CNOT)_{(n+1,1)}$ gate,$(CNOT)_{(1,n+1)}$ gate to the left of $I_2\otimes \Pi\mathsf{T}^e_{n,x}$. Add another $(CNOT)_{(1,n+1)}$ gate to the right of $I_2\otimes \Pi\mathsf{T}^e_{n,x}$. 
\State End If
\EndIf
\State End For
\EndFor
\State End
\end{algorithmic}
\end{algorithm}

Now we give Algorithm 6, combining all the Algorithms 2-5 for the generation of $(n+1)$-qubit circuit from $n$-qubit circuit.

\begin{small}
 \begin{algorithm}
\caption{Creating a $(n+1)$-qubit circuit to approximate any $U\in \Sf\Uf(2^{n+1})$ from a $n$-qubit circuit that approximates any $\hat{U}\in SU(2^{n})$}
\textbf{Provided:} $CNOT$ gates and 1 qubit rotation gates.\\
\textbf{Input:} $n$-qubit circuit that approximates any $\hat{U}\in \Sf\Uf(2^{n})$ and of the form mentioned in equation \ref{alg3} i.e. $\zeta(\Theta_{\zeta})\Psi(\Theta_{\psi})\psi(\Theta_{\psi})$ where all the terms have been defined in equation \ref{alg3}\\
\textbf{Output:}$(n+1)$-qubit circuit that approximates any ${U}\in \Sf\Uf(2^{n+1})$ \\
\begin{algorithmic}
\Procedure{}{}       \Comment{}
\State Add a qubit layer at the top/beginning of the circuit.
\State Create product of all $2^{n+1}$ special unitary diagonal matrices from product of all $2^{n}$ special unitary diagonal matrices using $\xi(F_i(R_z),F_i(R_z)),1\leq i\leq n$ in Algorithm \ref{multialgoz}.  
 \\

\For{$y=1:2^{n}-1;y++$}
\State Use Algorithm \ref{Createeven} create $\Pi\mathsf{T}^e_{n+1,y}$ using the function $\eta(y,2^{n+1},even)$
\State \State Use Algorithm \ref{Createodd} create $\Pi\mathsf{T}^o_{n+1,y}$ using the function $\eta(y,2^{n+1},odd)$
\State Add CNOT gates to convert $\Pi\mathsf{T}^o_{n,y}\rightarrow \Pi\mathsf{T}^o_{n+1,y}$\\ 
\State\textbf{End}
\EndFor
\State $\zeta(\Theta_{\zeta})\rightarrow \Pi_{i=1}^{(2^{n+1}-1)}\exp{(i \theta_{a}\chi^{-1}_{n+1}(a))}$, (see definition of $\chi$ at equation \ref{definition2})
\State Create a $(n+1)$-qubit $M_{n+1}ZYZ$ matrix $M^e_{y}$ from a $n$ qubit $M_nZYZ$ matrix using $\xi(F_n(R_z),F_n(R_z)),\xi(F_n(R_y),F_n(R_y))$ in Algorithm \ref{multialgoz} and Algorithm \ref{multialgoy}\\
\For{$y=1:2^{n}-1;i++$}
\State Create a $(n+1)$-qubit $M_{n+1}ZYZ$ matrix $M^e_{y}$ from a $n$ qubit $M_nZYZ$ matrix using $\xi(F_n(R_z),F_n(R_z)),\xi(F_n(R_y),F_n(R_y))$ in Algorithm \ref{multialgoz} and Algorithm \ref{multialgoy}\\
\State  Create a $(n+1)$-qubit block diagonal special unitary matrix $M^o_{y}$ from a $n$ qubit block diagonal special unitary matrix using $\xi(F_i(R_z),F_i(R_z)),1\leq i\leq n$ in Algorithm \ref{multialgoz}. and  $\xi(F_n(R_y),F_n(R_y))$ in Algorithm \ref{multialgoy}
\State\textbf{End}
\EndFor
\For{$y=1:2^n-1:y++$}
\State $\psi(\Theta_{\psi})\rightarrow M^e_0\Pi\mathsf{T}^e_{n+1,y} M^e_y {P}_{(y,2^{n+1},even)}$\\
\State $\psi(\Theta_{\psi})\rightarrow \psi(\Theta_{\psi})$\\
\State $\Phi(\Theta_{\phi})\rightarrow \Pi\mathsf{T}^o_{n+1,x} M^o_x \Pi\mathsf{T}^o_{n+1,x}$\\
\State $\Phi(\Theta_{\phi})\rightarrow \Phi(\Theta_{\phi})$
\State\textbf{End}
\EndFor
\State  $\zeta(\Theta_{\zeta})\psi(\Theta_{\psi})\Phi(\Theta_{\phi})$
\State \textbf{End Procedure}
\EndProcedure
\end{algorithmic}
\end{algorithm}
\end{small}

\begin{theorem}\label{count}
The circuit implementation of a special unitary matrix on $n$-qubits with $L$ layers using Algorithm $2$  requires at most $L(2.4^n+(n-5)2^{n})$ CNOT gates, $L({\frac{3}{2}\cdot} 4^n-\frac{5}{2}2^n+1)$ $R_z$ gates where $L$ is the number of iterations/layers.
\end{theorem}
\pf The proof follows from the above Algorithms. \hfill{$\square$}

\section*{Acknowledgment}

The authors thank Sabyasachi Chakraborty for his help with the numerical simulations using PARAM Shakti HPC at IIT Kharagpur. Rohit Sarma Sarkar acknowledges support through Prime Minister Research Fellowship (PMRF), Government of India.


\begin{tiny}
\begin{equation}\label{blkdiagcircuit}  
\Qcircuit @C=1em @R=.7em {
    &\lstick{1}&\gate{R_z}&\gate{}&\gate{}&\gate{}& \gate{}&\gate{}&\gate{}&\gate{}&\gate{}&\gate{}&\gate{R_z}\\
    &\lstick{2}&\qw&\gate{F_2(R_z)}\qwx[-1]&\gate{}\qwx[-1]\qwx[-1]&\gate{}\qwx[-1]& \gate{}\qwx[-1]&\gate{}\qwx[-1]&\gate{}\qwx[-1]&\gate{}\qwx[-1]&\gate{}\qwx[-1]&\gate{F_2(R_z)}\qwx[-1]&\qw\\
    &\lstick{\vdots}&\qw&\qw&\gate{\vdots F_i(R_z),3\leq i\leq n-2}\qwx[-1]&\gate{\vdots}\qwx[-1]& \gate{\vdots}\qwx[-1]&\gate{\vdots}\qwx[-1]&\gate{\vdots}\qwx[-1]&\gate{\vdots}\qwx[-1]&\gate{\vdots F_i(R_z),3\leq i\leq n-2}\qwx[-1]&\qw&\qw\\
    &\lstick{n-1}&\qw&\qw&\qw&\gate{F_{n-1}(R_z)}\qwx[-1]& \gate{}\qwx[-1]&\gate{}\qwx[-1]&\gate{}\qwx[-1]&\gate{F_{n-1}(R_z)}\qwx[-1]&\qw&\qw&\qw\\
    &\lstick{n}&\qw&\qw&\qw&\qw& \gate{F_n(R_z)}\qwx[-1]&\gate{F_n(R_y})\qwx[-1]&\gate{F_n(R_z)}\qwx[-1]&\qw&\qw&\qw&\qw\\}
\end{equation}
\end{tiny}



\end{document}